\documentclass[reprint,showpacs,preprintnumbers,amsmath,amssymb,prl]{revtex4-2}
\usepackage{amsmath}
\usepackage{amsfonts}
\usepackage{amsthm}
\usepackage{dsfont}
\usepackage{graphics}
\usepackage{graphicx,bm}
\usepackage[caption=false]{subfig}
\usepackage{amssymb}
\usepackage{mathrsfs}
\usepackage{braket}
\usepackage{float}
\usepackage{color}
\usepackage{url}
\usepackage[abs]{overpic}
\usepackage[usenames,dvipsnames]{xcolor}
\usepackage{commath}
\usepackage{subfig}
\usepackage{multirow}
\usepackage{verbatim}
\usepackage{array}

\begin{document}

\title{Space-Time from quantum Physics}

\author{Fabrice Debbasch}
\affiliation{Sorbonne Universit\'e, Observatoire de Paris, Universit\'e PSL, CNRS, LERMA, F-75005, Paris, France}, 

\date{\today}
\begin{abstract}
A construction of the real 4D Minkowski space-time starting from quantum harmonic oscillators is proposed. 
%The complex 4D Minkovski vector space $V$ is constructed first, through 
First, a 2D spinor space and its dual are derived from the standard commutation relations obeyed by the ladder operators of two independent 1D harmonic oscillators. The complex 4D Minkowski vector space $V$ is then constructed from these spinor space. The flat, real 4D Minkowski manifold is finally built as an approximate description of a manifold of unitary operators constructed from $V$. Lorentz invariance is recovered and several possible extensions are discussed, which connections to quantum optics and condensed matter physics.
%An extension to include curvature is also proposed.

\end{abstract}
%\end{abstract}
%\pacs{03.67.-a, 47.37.+q, 47.40.-x, 67.10.-j}

%\keywords{Spinors, Lorentz group, Lorentzian geometry, Quantum harmonic oscillator, Ladder operators}
\maketitle

%\section{Introduction}

The first quarter of the twentieth century witnessed several breakthroughs in physics. Special relativity, which was proposed by Einstein in 1905 \cite{Einstein05b} and later extended in 1915 into the relativistic theory of gravitation known as general relativity \cite{E15a, W84a}. Simultaneously, quantum physics, which also originated with Einstein in 1905 \cite{Einstein05a}, was developed first as a non relativistic theory of particles, which culminated in the equation proposed by Schr\"odinger in 1926 \cite{Schrodinger1926} and now bears his name. This equation was rapidly extended to the special relativistic realm. For example, the Dirac equation was introduced in 1928 \cite{Dirac28a} as a relativistic equation obeyed by the wave-functions of spin 1/2 particles. Quantum physics then morphed into a quantum theory of fields \cite{IZ12a}, making it possible to describe systems where particle numbers were themselves dynamical variables. And it is quantum field theory which, despite renormalisation issues, is today the natural framework used to describe the electro-weak and the strong interactions \cite{Quigg14a}. 

Despite all these achievements, modern physics has not yet been able to incorporate gravity into the quantum framework and develop a consistent quantum theory of gravitation. Several, apparently very different paths towards such a theory have been proposed \cite{K07a} and are still today active areas of research, but none has delivered a quantum gravity yet.

It seems today that building a quantum theory of gravity will require solving a lot of apparently different, but all very serious conceptual and technical/mathematical problems. One of them, perhaps the most conceptual one, is to reconcile the idea of classical space-time with what we know of quantum theory. Indeed, mathematically speaking, space-time is a differential Lorentzian manifold \cite{DNF84a} and general relativity thus relies heavily on geometry and analysis, while quantum physics, at its core, seems definitely more algebraic. One option is to take this as a fact, and try and build quantum gravity as a standard algebraic quantum theory taking place in a given geometric object called space-time. But this point of view seems somehow unnatural and many physicists since Pauli \cite{Pauli80a} have believed that space-time is not a fundamental concept and that it should be derived from quantum theory.

The aim of this Letter is to prove that the flat 4D Minkowski space-time of special relativity can indeed be seen as a local approximation of a real manifold which arises naturally from the algebra obeyed by the ladder operators \cite{Fey72a} of two independent `abstract' quantum harmonic oscillators.
% and that this construction can be extended to incorporate curvature. 
 This Letter does {\sl not} claim that the proposed construction is the only possible one, nor that it is the physically correct one, though it may be. The sole aim of this work is to show 
%that classical space-times can be 
that Minkowski space-time can be derived from a purely quantum framework. In other words, not only is the playground of special relativistic physics not alien to quantum physics, but it can be derived from it. If other, possibly more physically relevant constructions exist, is another question not dealt with in this work. 

We first show how to construct an abstract spinor space \cite{PR84a,Stew09a} and its dual from the algebra obeyed by the ladder operators of two independent`abstract' quantum harmonic oscillators. We then review how this abstract 
spinor space can be used with its dual to build a $4$D Minkowski vector space. 
We then elaborate on Lorentz invariance and show in particular that the space of the linear transformations which leave invariant the algebra obeyed by the ladder operators and which do not mix spinors and dual spinors is simply the Lorentz group. 
We then discuss the differences between the Minkowski vector space introduced earlier and the usual $4$D space-time manifold
of special relativistic physics and finally offer a local construction of the Minkowski space-time space-time manifold. 
The Letter concludes with a summary and a discussion of all results, with special emphasis on possible extensions. 
%(Section 6 offers a summary and a careful discussion of the construction presented in this article.)

%\section{Construction of the abstract spinor space}

Consider a Hilbert space $\mathcal H$ and two linearly independent operators $a$ and $b$ defined on $\mathcal H$ which obey the algebra: 
\begin{eqnarray}
\left[ a, a^\dagger \right] & = & \left[ b, b^\dagger \right]  = 1 \nonumber\\
%\left[ b, b^\dagger \right] & = & 1 \nonumber\\
\left[a, b \right] & = & \left[a, b^\dagger \right] = 0.
\label{eq:laddercom}
%\nonumber\\
%\left[a, b^\dagger \right] & = & 0.
\end{eqnarray}
This algebra can be realized, for example, by combining two independent harmonic oscillators, choosing as $\mathcal H$ the tensor
product of their Hilbert spaces and by retaining as operators $a$, $a^\dagger$, $b$, $b^\dagger$ the standard ladder operators of the two oscillators. 

The operators $a$ and $b$ can be used to build the two operators
\begin{eqnarray}
\beta_0 & = & \frac{1}{\sqrt{2}} \left( a + i b^\dagger \right)\nonumber\\
\beta_1  & = & \frac{1}{\sqrt{2}} \left(  a^\dagger + i b \right).
\end{eqnarray}
The algebra (1) obeyed by $a$, $a^\dagger$, $b$ and $b^\dagger$ is equivalent to the algebra:
\begin{eqnarray}
\left[\beta_0, \beta_1\right]  & = & 1 \nonumber\\
\left[\beta_0, \beta_0^\dagger \right]  =  \left[\beta_1, \beta_1^\dagger \right] & = & \left[\beta_0, \beta_1^\dagger \right] = 0.
\label{eq:betacom}
% \nonumber\\
%\left[\beta_1, \beta_1^\dagger \right] & = & 0  \nonumber\\
%\left[\beta_0, \beta_1^\dagger \right] & = & 0 . 
\end{eqnarray}

The commutator defines a bilinear antisymmetric form on the space spanned by $(\beta_0, \beta_1, \beta_0^\dagger, \beta_1^\dagger)$. The above commutation relations show that this form is degenerate. There are however two sub-spaces on which the form is not degenerate, and these are the sub-space $S$ spanned by $(\beta_0, \beta_1)$ and the sub-space $\bar S$ spanned by $(\beta_0^\dagger, \beta_1^\dagger)$. Each of these sub-spaces is a so-called abstract spinor space \cite{PR84a,Stew09a}.

By definition, a spinor $\sigma$ in $S$ can be decomposed on the basis $\beta$, and we write
%\begin{eqnarray}
$\sigma  =  \sum_{A = 0}^1 \sigma^A \beta_A$
%\end{eqnarray}
or, introducing Einstein's summation convention, $\sigma = \sigma^A \beta_A$. 
As mentioned above, the commutator is a non-degenerate, bilinear, anti-symmetric form on this space. To make writing with components easier, we denote the commutator by $\epsilon$.
Its components $(\epsilon_{AB})$ in the basis $\beta$ are $\epsilon_{01} = - \epsilon_{10} = 1$ and $\epsilon_{00} = \epsilon_{11} = 0$. 

The two operators $(\beta^\dagger_1, \beta^\dagger_0)$ span another spinor space ${\bar S}$, called the dual spinor space. To be consistent with the literature on abstract spinors, we introduce the notation ${\bar \beta} = ({\bar \beta}_{\bar 0} = \beta^\dagger_1, {\bar \beta}_{\bar 1} = \beta^\dagger_0)$. Components of a spinor ${\bar \sigma}$ in the dual spinor space will be denoted by ${\bar \sigma}_{\bar A}$ and the Poisson bracket, as a bilinear anti-symmetric form on ${\bar S}$ is denoted by ${\bar \epsilon}$, with components
${\bar \epsilon}_{{\bar A} {\bar B}}$.

%\section{Construction of Minkovski vector space}

Let us now introduce the operators
\begin{eqnarray}
{\gamma}_{{\bar 0} 0} & = & {\bar \beta}_{\bar 0} \beta_0 \nonumber \\
{\gamma}_{{\bar 1} 1}& = & {\bar \beta}_{\bar 1} \beta_1  \nonumber \\
{\gamma}_{{\bar 0} 1} & = & {\bar \beta}_{\bar 0} \beta_1  \nonumber \\
{\gamma}_{{\bar 1} 0}& = & {\bar \beta}_{\bar 1} \beta_0 . 
\end{eqnarray}
The family $(\bar \beta)$ spans a $4D$ complex vector space $V$ and one can write any vector $v \in V$ as 
$V = v^{{\bar A}B} {\gamma}_{{\bar A} B}$. Note that, at this stage, the 4 indices are not $0, 1, 2, 3$, but ${\bar 0} 0, {\bar 1} 1, {\bar 0} 1, {\bar 1} 0$.

The commutator induces in $V$ a metric $\eta$ defined by $\eta_{({\bar A}A) ({\bar B}B)} = {\bar \epsilon}_{{\bar A}{\bar B}} \epsilon_{AB}$
and we denote the corresponding scalar product by a dot.
A direct computation shows that the only non vanishing scalar products between the $\gamma$'s are $\gamma_{{\bar 0}0} \cdot \gamma_{{\bar 1}1} =  
- \gamma_{{\bar 0}1} \cdot \gamma_{{\bar 1}0} =  1$. In particular, each $\gamma$ has a vanishing scalar product with itself and is thus a null vector. 
Also, replacing in the definitions of the $\gamma$'s the $\bar \beta$ by their expressions in terms of the $\beta^\dagger$'s, one finds
${\gamma}_{{\bar 0} 0}  =  {\beta}_{1}^\dagger \beta_0$, ${\gamma}_{{\bar 1} 1} =  {\beta}_{0}^\dagger \beta_1$,
${\gamma}_{{\bar 0} 1}  =  {\beta}_{1}^\dagger \beta_1$, ${\gamma}_{{\bar 1} 0} =  {\beta}_{0}^\dagger \beta_0$.
%\begin{eqnarray}
%{\gamma}_{{\bar 0} 0} & = & {\beta}_{1}^\dagger \beta_0 \nonumber \\
%{\gamma}_{{\bar 1} 1}& = & {\beta}_{0}^\dagger \beta_1  \nonumber \\
%{\gamma}_{{\bar 0} 1} & = & {\beta}_{1}^\dagger \beta_1  \nonumber \\
%{\gamma}_{{\bar 1} 0}& = & {\beta}_{0}^\dagger \beta_0 . 
%\end{eqnarray}
This shows that ${\gamma}_{{\bar 0} 1}$ and ${\gamma}_{{\bar 1} 0}$ are self-adjoint while ${\gamma}_{{\bar 0} 0}$ and ${\gamma}_{{\bar 1} 1}$ are dual to each other. 
Thus, the set of $\gamma$'s is the equivalent of what is called a null $4$-bein \cite{PR84a, W84a, Y08} in Lorentzian geometry. 
Another $4$-bein is
\begin{eqnarray}
e_0  & = & \frac{1}{\sqrt{2}}\, \left({\gamma}_{{\bar 0} 1}  +  {\gamma}_{{\bar 1} 0} \right) \nonumber \\
e_1  & = & \frac{1}{\sqrt{2}}\, \left({\gamma}_{{\bar 0} 1} -  {\gamma}_{{\bar 1} 0} \right) \nonumber \\
e_2  & = & \frac{1}{\sqrt{2}}\, \left({\gamma}_{{\bar 0} 0}  +  {\gamma}_{{\bar 1} 1} \right) \nonumber \\
e_3  & = & \frac{1}{i \sqrt{2}}\, \left({\gamma}_{{\bar 0} 0}  -  {\gamma}_{{\bar 1} 1} \right).
\end{eqnarray}

The components of the metric $\eta$ in the $e$-basis read $(\eta_{\mu \nu}) = \mbox{diag}(-1, 1, 1, 1)$, which is the standard form of Minkowski metric. Also, all four $e$ vectors are Hermitian operators. But this does not make the space $V$ identical to the physical 4D Minkowski space-time. 
%Section @@@ will review the differences and construct from $V$ the 
%standard 4D Minkovski space-time manifold, together with the usual, physical 4D Minkovski vector space. 
Before constructing from Minkowski vector space the Lorentzian space-time manifold, let us discuss first how Lorentz invariance emerges in the present, operator oriented context.

%\section{Lorentz invariance} 

It is natural to wonder if there are linear transformations in $S \cup {\bar S}$ which leave the original commutation relations (\ref{eq:laddercom}) obeyed by $(a, a^\dagger, b, b^\dagger)$ or, equivalently, the commutation relations obeyed by $(\beta_0, \beta_1, \beta_0^\dagger, \beta_1^\dagger)$ invariant. Any linear transformation in $S$ induces a linear, dual transformation $\bar S$ and, thus, a linear transformation in $S \cup {\bar S}$. So, are there for example linear transformations in $S$ which, together with their dual, leave the commutation relations invariant? 

Without loss of generality, an arbitrary linear transformation in $S$ can be written as
\begin{eqnarray}
\beta_0' & = & p \beta_0 + i q \beta_1 \nonumber\\
\beta_1' & = & i r \beta_0 + s \beta_1.
\end{eqnarray}
where $p$, $q$, $r$, $s$ are four arbitrary complex numbers.
It induces in $\bar S$ the transformation:
\begin{eqnarray}
(\beta_0^\dagger)' & = & {\bar p} \beta_0^\dagger - i {\bar q} \beta_1^\dagger \nonumber\\
(\beta_1^\dagger)' & = & - i {\bar r} \beta_0^\dagger + {\bar s} \beta_1^\dagger
\end{eqnarray}
where bars over complex numbers denote complex conjugation. The last three commutation relations obeyed by $(\beta_0, \beta_1, \beta_0^\dagger, \beta_1^\dagger)$ in equation (\ref{eq:betacom}) 
are trivially invariant under the above transformations and the invariance of the first commutation relation is equivalent to $p s + q r = 1$. The four complex numbers $p$, $q$, $r$, $s$ are thus restricted by a single (complex) relation, and we are thus dealing with a family of transformations which depend {\sl a priori} on $3$ complex, or equivalently $6$ real parameters. We will now show that these transformations coincide with the Lorentz transformations. 

As a preliminary, we first show that any complex number, say $u$, can be written as the squared cosine of another complex number, say $\theta$. 
The equation $u = \cos^2 \theta$
transcribes into $v = 2u - 1 =  \cos(2 \theta)$. Introducing $x = \exp(2 i \theta)$ leads to $x^2 - 2v x + 1 = 0$, which admits two (possibly identical) complex solutions. One can thus always find an $x$ which solves the problem. Writing then $x = \mid x \mid \exp(i \phi_x)$ with $\phi_x \in (0, 2 \pi)$ and $\theta = \theta_r + i \theta_i$, with $(\theta_r, \theta_i) \in 
{\mathbf R}^2$, the equation $x = \exp(2 i \theta)$  for $\theta$ can be solved by choosing for example $\theta_r = \phi_x/2$ and $\theta_i = - (\ln \mid x \mid) / 2$ (note that $x$ does not vanish for any value of $u$). 

Coming back to the relation $p s + q r = 1$, 
we iIntroduce a complex $\theta$ such that $p s = \cos^2 \theta$. 
%(see Supplemental Material). 
This implies that $q r = \sin^2 \theta$. One can then introduce two complex numbers $\xi$ and $\zeta$ such that $p = \exp( i \xi) \cos \theta$ and $i r = \exp(- i \zeta) \sin \theta$.  Since $p s = \cos^2 \theta$ and $q r = \sin^2 \theta$, one gets immediately
$s = \exp(- i \xi) \cos \theta$ and $i q = - \exp(- i \zeta) \sin \theta$. 

A straightforward computation shows that, conversely, any linear transformation of $S$ represented in the $(\beta_0, \beta_1)$ basis by a matrix $L$ of the form
\begin{equation}
L = \begin{pmatrix}
\exp( i \xi) \cos \theta & - \exp(i \zeta) \sin \theta\\
\exp( -i \zeta) \sin \theta &  \exp(-i \xi) \cos \theta\\
\end{pmatrix}
\end{equation}
with arbitrary complex $\theta$, $\xi$ and $\zeta$ preserves the commutation relations. The matrix $L$ can be rewritten as the exponential of a complex linear combination of the three Pauli matrices and is thus identical to the action of an arbitrary Lorentz transformation on $2$-spinors \cite{ED84a,IZ12a}. 

The Lorentz transformations are thus the only linear transformations of $S \cup {\bar S}$ which, not only preserve the commutation relations, but also leave each of the subspaces $S$ and $\bar S$ invariant (as sets, not point-wise). 

One final remark is in order. 
%The action of the Lorentz transformations on the ladder operators reads: 
%
Writing the Lorentz transformations in terms of the original ladder operator, one finds:
\begin{eqnarray}
a' + i (b^\dagger)' & = & p \left[ a + i (b^\dagger) \right] + i q \left[ a^\dagger + i b \right] \nonumber \\
(a^\dagger)' + i b' & = & i r \left[ a + i (b^\dagger) \right] + s \left[ a^\dagger + i b \right].
\end{eqnarray}
Taking the dual of these equations, one obtains:
\begin{eqnarray}
(a^\dagger)' - i b' & = & {\bar p} \left[ a^\dagger - i b \right] - i {\bar q} \left[ a - i b^\dagger \right] \nonumber \\
a' - i (b^\dagger)' & = & - i  {\bar r} \left[ a^\dagger - i b \right] + {\bar s} \left[ a - i b^\dagger \right].
\end{eqnarray}
These four equations can be combined to deliver:
\begin{eqnarray}
& & \hspace{-0.4cm}a'   =  \frac{1}{2} \left[ ( p + {\bar s}) a - (q + {\bar r}) b + i (q - {\bar r}) a^\dagger + i (p - {\bar s}) b^\dagger 
\right] \nonumber\\
& & \hspace{-0.4cm}b'  =  \frac{1}{2} \left[ ( r + {\bar q}) a + (s + {\bar p}) b - i (s - {\bar p}) a^\dagger + i (r - {\bar q}) b^\dagger 
\right].
\end{eqnarray}
If one interprets $a$ and $b$ to be ladder operators for two independent harmonic oscillators, a Lorentz transformation actually mixes these two oscillators, and also mixes their creation and destruction operators, to generate two new oscillators still independent of each other.  

%\section{Minkovski space-time manifold}

%\subsection{The problem}

The 4D space $V$ exhibits several key differences with the 4D Minkowski space-time. First, $V$ is a vector space, and not a manifold. Second, $V$, as built above, is a complex vector space. One can naturally argue that the physical Minkowski vector space, which is tangent to the space-time manifold, is a subspace of $V$, but then, why does physics only deal with that subspace? Or is the physical Minkowski space-time actually complex? The third difference is more subtle. By construction, all elements of $V$ are operators in a space on which a bilinear antisymmetric form, which we have called the commutator, is defined. One can therefore compute the commutator of different elements in $V$ and, in particular, the commutator of the $e_a$'s with each other. In practice these commutators can be found from the commutation relations between the $\gamma$'s, which can be derived from those of the $\beta$'s and $\beta^\dagger$'s. One find that the $\gamma$'s do not commute with each other, and neither do the $e$'s. Also, all commutators between the $e$' are quadratic in the ladder operators $a$, $a^\dagger$, $b$, $b^\dagger$. For example, $\left[ e_0, e_1 \right] = -i \left(a^\dagger b^\dagger + a b \right)$.

This might have been expected and shows that neither the $\gamma$'s nor the $e$'s are identical to a $4$-bein in usual Minkowski vector space. One might be tempted to interpret the non commutation of the $4$-bein in terms of curvature, but there is no manifold at this stage of the computation. Moreover, it is easy to check that the commutator $[e_0, e_1]$ actually lies outside of $V$. In particular, the $e$'s are Hermitian, and their commutators are therefore
anti-Hermitian, so outside of $V$. Observe also that the $e$'s are quadratic in the ladder operators $a$, $b$, $a^\dagger$, $b^\dagger$, and so are their commutators. 

To summarize the above discussion, what we have at this stage is a non-commutative 4D complex Minkowski vector space embedded in a larger operator space. The question is: can we build from that the usual arena of physics {\sl i.e.} a real 4D Lorentzian manifold? 

%\subsection{The solution}

%\subsubsection{Flat 4D Minkovski space-time}

The 
$4$D Minkowski manifold is flat and can thus be viewed as an affine space, and there is a standard way to construct affine submanifolds of a vector space. 
Suppose you want to build an affine space from $V$ defined above. Consider the space $W$ of all operators acting on the Hilbert space $\mathcal H$ and pick up an operator $u$ in $W$ which is not in $V$. The space $u + V = \left\{ u + v, v \in V \right\}$ is then a flat submanifold of $W$ with tangent $V$ at each point. Note that this construction cannot work if one starts from a vector space which is not embedded in a larger vector space. 

So, one first tentative way to obtain physical Minkowski $\mathcal M$ space-time would be to pick an arbitrary $u$ outside of $V$ and to identify $\mathcal M$ with the set of all
$u + \alpha^a e_a$, $a \in \left\{ 0, 1, 2, 3\right\}$ and $\alpha \in {\mathbb R}^4$. This however presents at least two shortcomings. First, the $4$-bein vectors in $V$ still do not commute, so the Minkowski manifold $u + V$ is not the standard one. Second, the restriction to real $\alpha$'s comes out of nowhere in this construction. 

These two shortcomings have a common solution
Instead of considering operators
of the form $u + \alpha^a e_a$ with $\alpha \in {\mathbb R}^4$, consider the operators $M(u, \alpha) = u \exp\left ( i \alpha^a e_a\right)$ with $u$ unitary and not in $V$. Possible choices for
$u$ are $u = \exp(i \delta) {\mathcal I}$ where $\delta$ is an arbitrary real number and $\mathcal I$ is the identity operator, which is not in $V$. 
The operators $M(u, \alpha) = u \exp\left ( i \alpha^a e_a\right)$ form a submanifold ${\bar {\mathcal M}}$ of $W$. 
Since the $e$'s are Hermitian, an operator in ${\bar {\mathcal M}}$ is unitary iff its $4$ $\alpha$ coordinates are real. We denote the set of all unitary operators in ${\bar {\mathcal M}}$ by ${\bar {\mathcal M}}_U$.

Suppose now one looks only at points which are `close' to $u$, say points for which the four $\alpha$'s are at the most $O(\epsilon)$ with $0 < \epsilon \ll 1$. 
It is natural, for these points, to introduce the rescaled coordinates $({\bar \alpha})$ defined by $\alpha^a = \epsilon\,  {\bar \alpha}^a$ for $a = 0, 1, 2, 3$. With this definition, all ${\bar \alpha}$'s can reach $O(1)$.
One can then expand
$M(u, \epsilon {\bar \alpha)}$ at first order in $\epsilon$ around $\epsilon = 0$ and obtain
%\begin{equation}
$M(u, \epsilon {\bar \alpha)} = u \left( 1 + i \epsilon{\bar \alpha^a} e_a + O(\epsilon^2) \right)$.
%\end{equation}
We now introduce the rescaled $4$-bein ${\bar e}_a = \epsilon\,  e_a$, and write
%so a point $M(u, \epsilon {\bar \alpha})$ now reads
\begin{equation}
M(u, {\bar \alpha}) = u \left( 1 + i {\bar \alpha^a} {\bar e_a} + O(\epsilon^2) \right).
\label{eq:Mepsilon}
\end{equation}
Since the $e$'s are quadratic in the ladder operators, rescaling the $e's$ by $\epsilon$ is tantamount to rescaling the ladder operators by
$\sqrt{\epsilon}$. The rescaled commutation relations between the $4$-bein vectors thus take the form
$\left[ {\bar e}_a, {\bar e}_b \right] = O(\epsilon) C_{ab}$ where the $C_{ab}$'s are quadratic in the rescaled ladder operators. This means that the commutators of the rescaled $4$-bein vectors tend to zero with $\epsilon$. Thus, close to $u$, ${\bar {\mathcal M}}_U$ looks like the standard, real Minkowski manifold.
This proves that the usual flat space-time of special relativity can be recovered from standard quantum theory. 
Note that introducing the new $4$-bein $E_a = i {\bar e_a}$ makes it possible to write (\ref{eq:Mepsilon}) as
%\begin{equation}
$M(u, {\bar \alpha}) = u \left( 1 + {\bar \alpha^a} {E_a} + O(\epsilon^2) \right)$,
%\end{equation}
where the $i$ factor does not explicitly appears.

In the above construction, a portion of the flat Minkowski space-time appears as the portion of the affine space $- i + V$ for which all $\alpha$'s at at most $O(1)$. Suppose now a physicist is using Lorentzian `physical' coordinates $(x^a)$, a = 0, ..., 3 and works in (or has access to) a region of Minkowski space-time of size $L$ is these coordinates. This region can be described in the above framework by assuming that $x = {\bar L} {\bar \alpha}$ with ${\bar L} \ge L$.

%\section{Conclusion}

%\subsection{Summary}

We have shown that the real $4$D Minkowski space-time can be constructed locally from the ladder operators of two independent 1D quantum oscillators. The ladder operators generate an abstract spinor space and its dual which can in turn be used to build the complex $4$D Minkowski vector space. Part of this vector space generates a space of unitary operator which, locally, looks like the standard real $4$D 
Minkowski space-time.

%\subsection{Discussion}

Let us now discuss this result, focusing on possible extensions. 
From the quantum point of view, the procedure described in this Letter is but a special case of a very general problem. 
Consider a collection of $N$ independent `abstract' 1D quantum harmonic oscillators {\sl i.e.} $N$ independent quantum systems characterized by ladder operators obeying the standard commutation relations. Here, `abstract' means that there is no space-time available at this stage and that the ladder operators have  therefore no relations with physical position and momentum operators.
Then ask yourself what are the unitary operators or, equivalently, the Hermitian operators which can be built from these independent oscillators {\sl i.e.} from their ladder operators. 
%The logarithm of a unitary operator is anti-Hermitiian. Since the product of any anti-Hermitian operator by the complex number $i$ delivers a Hermitian operator, the original question comes down to finding the Hermitian operators which can be built from the ladder operators. 
This question is completely natural from the point of view of quantum physics, it is interesting {\sl per se}, it connects with several domains, including quantum optics and condensed matter physics, but it does not seem to have much to say about space-time. 
This Letter shows that, contrary to what one might think {\sl prima facie}, the case $N = 2$ delivers Hermitian operators quadratic in the ladder operators which, in turn, deliver a manifold of unitary operators which, locally, looks like the usual, real $4$D Minkowski space-time. 

But what about other values of $N$? The case $N = 1$ is nearly trivial and does not seem to connect to space-time physics (computations not shown). Other values of $N$ should be investigated to determine if and what manifolds they generate, taking into account polynomials of arbitrary degrees in the ladder operators. For example, do higher values of $N$ and/or polynomial of higher degrees deliver Lorentzian space-times of higher dimensions? Or space-times with more than a single time-coordinate? 

As constructed in this Letter, flat space-time is but a local approximation of a manifold of non-commutating operators. The above problem involving $N$ independent quantum harmonic oscillators thus has connections with non-commutative geometry \cite{CCS23a}, and this should be investigated.

From the classical point of view, the most natural question is about general relativity. Can the above procedure be extended to deliver curved
Lorentzian manifolds? Could this be done by allowing for example the oscillators to interact with each other? And could this pave the way to a 
possible laboratory quantum simulations of general relativistic space-times {\sl i.e.} of relativistic gravitation, for example in the contexts of quantum optics or condensed matter physics? 

Finally, one can only wonder if and how matter fits into the picture developed here. The link between space-time and quantum harmonic oscillators presented in this article seems to suggest that matter and space-time may be two sides of the same coin. If so, what is exactly that coin, and how does what we call dynamics emerge from a unified quantum picture?

\section{Supplemental material}

%\subsection{Alternative expressions of the $4$-bein vectors $e_a$}

The $4$-bein vectors $e_a$, $a = 0, ..., 3$ can be expressed in terms of the ladder operators. One finds:
\begin{eqnarray}
e_0  & = & \frac{1}{2\sqrt{2}}\, \left(a a^\dagger + a^\dagger a + b b^\dagger + b^\dagger b \right) \nonumber \\
e_1  & = & \frac{i}{\sqrt{2}}\, \left(a b - b^\dagger a^\dagger \right) \nonumber \\
e_2  & = & \frac{1}{2 \sqrt{2}}\, \left(a^2 + (a^\dagger)^2 + b^2 + (b^\dagger)^2 \right) \nonumber \\
e_3  & = & - \frac{i}{2 \sqrt{2}}\, \left(a^2 - (a^\dagger)^2 - b^2 + (b^\dagger)^2 \right).
\end{eqnarray}
It is interesting to introduce the hermitian `position' operators $X_a$, $X_b$ and `momentum' operators $P_a$, $P_b$ of the two original independent oscillators, which we define by:
%\begin{eqnarray}
%a & = & \mu_a^{1/2} \left( X_a + i \mu_a^{-1} P_a \right)\nonumber\\
%b & = & \mu_b^{1/2} \left( X_b + i \mu_b^{-1} P_b \right).
%\end{eqnarray}
\begin{eqnarray}
a & = &   X_a + i P_a \nonumber\\
b & = &  X_b + i  P_b .
\end{eqnarray}
%If the oscillator were real, physical harmonic oscillators in space-time, the coefficients $\mu$ would represent the products of the `mass' of each oscillator by its `frequency'. But, since there is no space-time at this stage, the $\mu$'s only track the fact that the two oscillators may be `different. 
The commutation relations obeyed by the $P$'s and $X$'s are $\left[ P_{a/b}, X_{a/b} \right] = -i$, as they should. But remember 
that the $X$'s are not usual physical positions, which are denoted by $\alpha$ in this Letter. The $P$'s are not physical momenta either. 
Calling the $P$'s and $X$'s `momenta'  and `positions' is thus a pure matter of habit and convention and does not point to the usual physical meaning of these terms, but rather to the fact that their commutators are equal to $- i$. 

In terms of these harmonic oscillators position and momentum operators, the four $e$-vectors read:
%A direct computation delivers
\begin{eqnarray}
e_0  & = & \frac{1}{\sqrt{2}}\, \left(X_a^2 + P_a^2 + X_b^2 + P_b^2 \right) \nonumber \\
e_1  & = & - \sqrt{2}\, \left(P_a X_b + X_a P_b \right) \nonumber \\
e_2  & = & \frac{1}{\sqrt{2}}\, \left(X_a^2 - P_a^2 + X_b^2 - P_b^2 \right) \nonumber \\
e_3  & = & \sqrt{2}\, \left(P_a X_a - P_b X_b \right).
\end{eqnarray}
The first $4$-bein vector $e_0$ is thus the sum of the energies of the two oscillators. The other vectors do not have a comparable direct physical interpretation. Note however that
\begin{eqnarray}
e_0 + e_2 & = & \sqrt{2}\, \left(X_a^2 + X_b^2 \right) \nonumber \\
e_0 - e_2 & = &  \sqrt{2}\, \left(P_a^2 + P_b^2 \right).
\end{eqnarray}
These two vectors therefore represent respectively the total potential energy and the total kinetic energy of the two oscillators. 
One also find
\begin{eqnarray}
e_1 + i e_3 & = &- i  \sqrt{2}\, \left(P_b - i P_a \right)  \left(X_b - i X_a \right)  \nonumber \\
e_1 - i e_3 & = &+ i  \sqrt{2}\, \left(P_b + i P_a \right)  \left(X_b + i X_a \right),
\end{eqnarray}
which can be combined into 
\begin{eqnarray}
e_1 + i e_3 & = & - i \sqrt{2} P_Z Z \nonumber \\
e_1 - i e_3 & = & + i \sqrt{2} {\bar P}_Z {\bar Z}
\end{eqnarray}
where $Z = X_b - i X_a$ and $P_Z = P_b - i P_a$. The vectors $e_0 \pm e_2$ read simply
\begin{eqnarray}
e_0 + e_2 & = & \sqrt{2}\, \mid Z \mid^2 \nonumber \\
e_0 - e_2 & = &  \sqrt{2}\, \mid P_Z \mid^2.
\end{eqnarray}

\end{document}